\documentclass[preprint,nofootinbib]{revtex4-1}

\usepackage{amsmath}
\usepackage{amsfonts,amssymb}
\usepackage{hyperref}
\usepackage{graphicx}
\usepackage[utf8]{inputenc}
\usepackage{xcolor}
\usepackage{bm}

\hypersetup{
    colorlinks, 
    linktoc=all, 
    linkcolor=red,  
  citecolor=hyptxt,
  urlcolor=blue}
  \hypersetup{
  citebordercolor=,
  filebordercolor=green,
  linkbordercolor=blue
}
\definecolor{hyptxt}{rgb}{0.7, 0.4, 0.9}

\begin{document}

\title{Impact of inhomogeneities on slowly rolling Quintessence: implications for the local variations of the fine-structure constant}

\author{Leonardo Giani$^{1}$}
\email{uqlgiani@uq.edu.au}

\affiliation{$^1$ The University of Queensland, School of Mathematics and Physics, QLD 4072, Australia}

\author{Emmanuel Frion$^{2}$}
\email{emmanuel.frion@helsinki.fi}

\affiliation{$^2$ Helsinki Institute of Physics (HIP), P.O. Box 64, FIN-00014 University of Helsinki, Finland}

\author{Oliver F. Piattella$^{3}$}
\email{oliver.piattella@cosmo-ufes.org}

\affiliation{$^3$ Department of Physics, Universidade Federal do Esp\'irito Santo, Avenida Fernando Ferrari 514, 29075-910 Vit\'oria, Esp\'irito Santo, Brazil}



\begin{abstract}
We study how the evolution of a Dark Energy Quintessence fluid is modified by the presence of a matter inhomogeneity. To do so, we study linear perturbations of a flat FLRW background containing dust and a slowly rolling scalar field. Under the assumptions of spherical symmetry and a static density contrast, \textit{i.e.} $\dot{\delta}=0$, we obtain simple analytical solutions for perturbations in the matter and dark energy-dominated epochs. As a consequence, we show that perturbations of the scalar field, if a coupling \textit{à la} Bekenstein is assumed, trigger a spatial dependence of the fine-structure ``constant" which then varies as $\Delta \alpha \propto 1/r$. We finally highlight that such variations can be constrained with spectroscopic observations of stars from within our galaxy, therefore offering a new probe of the nature of Dark Energy.
\end{abstract}

\maketitle

\section{Introduction}
The discovery that the expansion of the universe is accelerating \cite{Riess:1998cb, Perlmutter:1998np} forced cosmologists to accept the idea that the dominant component of energy-momentum in the universe is an exotic fluid that exhibits negative pressure or is a form of anti-gravity. The lack of understanding of such a component is reflected in its label, Dark Energy (DE).
The most successful (and maybe natural) description of it is a cosmological constant $\Lambda$. Together with Cold Dark Matter (CDM), and in the framework of General Relativity, these components are the pillars of the concordance model of cosmology known as $\Lambda$CDM.
On the other hand neither $\Lambda$ nor CDM are completely satisfactory from the theoretical point of view, the former because of the difficulty in realising it as a model of vacuum energy, see for example Refs.~\cite{RevModPhys.61.1, MARTIN2012566}, the latter because a weakly interacting DM particle requires physics beyond the standard model of particle physics.

For the above reasons, in the last 20 years, the scientific community has explored the viability of alternative DE and DM candidates. For what concerns DE, in particular, several proposals were made in which $\Lambda$ is replaced by a new dynamical quantity. When such a quantity is described by a canonical scalar field $\varphi$ minimally coupled with gravity, with standard kinetic energy and potential $V(\varphi)$, it is usually referred to as \textit{Quintessence}, see for example Refs.~\cite{Wetterich:1987fm,2010deto.book.....A,Nojiri:2017ncd}.
The presence of a new degree of freedom naturally poses the question of how and if such a new component would interact with ordinary matter and could then potentially result in a violation of the equivalence principle, see for example Ref.~\cite{Uzan:2010pm}.
The possibility of such a violation through a dependence of the universal constants on their spacetime position  was already discussed by Dicke in 1959 \cite{dicke1959new}. 
Later on, in the 1980s, the specific case  of the fine-structure constant, $\alpha$, was discussed by Bekenstein in the pioneering work~\cite{Bekenstein:1982eu}. Bekenstein, at the time, concluded that tests of the equivalence principle rule out spacetime variability of $\alpha$ at any level.
In the last decades, however, huge improvements have been made from the experimental point of view, leading to tight constraints on the variation of $\alpha$, see for example Refs.~\cite{Chiba:2001er,Uzan:2010pm, FLAMBAUM_2007,Leal:2014yqa,Holanda:2015oda,Pinho_2016,Hees:2020gda}, and  claims of statistical evidences of $\alpha$ variations, see for example Refs.~\cite{Murphy_2003,Webb:2010hc}, as well as studies pointing otherwise, like \cite{Milakovic:2020tvq,Wilczynska:2020rxx}.  It is important to note that the connection between DE and the variation of $\alpha$ is of great importance from the observational point of view, as it is indeed possible to relate constraints on $ \Delta\alpha$ to constraints on DE parameters, see for example Refs.~\cite{Chiba:2003hz,Calabrese_2014,  Martins_2015, Martins2_2015}.
For the above reasons, the subject is nowadays very popular, since a violation of the equivalence principle could potentially confirm or rule out several alternative theories of gravity \cite{Tino:2020nla}. One could in general distinguish between variations on the value of $\alpha$ on large scales or on local scales. On cosmological scales, these could be motivated by a modification of the gravitational theory due to the dynamical behaviour of the DE field and were extensively studied in the literature, see for example Refs.~\cite{Olive:2001vz,Marra:2005yt,Barrow:2009nt, Barrow:2011kr,Barrow:2013uza,Sloan:2013wya,Graham:2014hva,vandeBruck:2015rma,Fritzsch:2016ewd}. On local scales, the variation of $\alpha$ is related to the local gravitational field, see for example Refs.~\cite{Bekenstein:2009fq,Shaw:2005gt, Shaw:2006zs, Barrow:2014vva,Barrow:1999qk,Barrow:2002db}. 

As discussed in Ref.~\cite{Mota:2003tm}, variations of $\alpha$ induced by matter inhomogeneities strongly depend on the specific way these are formed and the stage of their evolution.
Time variations of $\alpha$ induced by perturbations of a canonical scalar field due to inhomogeneities were studied analyzing solutions of the Klein-Gordon equation in Ref.~\cite{Barrow:2002db}, where it is shown that such time variations decay at any stage of the evolution of the Universe. Furthermore, in Ref.~\cite{Shaw:2005gt}, it was shown analytically for the McVittie and LTB geometries that constraints on the cosmological time variation of $\alpha$ obtained on Earth scales are reliable on larger scales. 
Spatial variations of $\alpha$ can be triggered by space-dependent scalar field perturbations. They have a quantifiable impact on the CMB, see for example Ref.~\cite{Smith:2018rnu}, but such an impact becomes very small if we consider large multipoles $\ell$, \textit{i.e.} if we consider variation of $\alpha$ on sufficiently small scales. Recently, in Ref.~\cite{Hees:2020gda}, constraints on the variation of $\alpha$ were obtained by spectroscopic measurements of stars in the proximity of the supermassive black hole in our galactic center. 
The possibility of testing local variations of $\alpha$ in regimes of strong gravitational potential motivates further investigations of the interplay between matter, gravitational potential and DE.

In this work, we study the behavior of scalar perturbations of a Quintessence DE fluid in the proximity of a spherically symmetric inhomogeneous perturbation of the FLRW background. We model these inhomogeneities with the goal of describing stable structures. 
We find simple analytical solutions for which the scalar field perturbations trigger a local variation of $\alpha$ which goes as $1/r$, where $r$ is the radial distance from the center of the structure, and decays with time. This result is potentially interesting for its implications on spatial variations of $\alpha$ within the gravitational potential surrounding a stable structure.

Throughout the paper, we use $c = M_P = 8\pi G_N =1$ units.


\section{Coupling of the scalar field and varying alpha}\label{sec2}

Following Ref.~\cite{Bekenstein:1982eu}, one can obtain a dynamical fine-structure constant $\alpha$ by considering a coupling $B_F(\varphi)$ between a cosmological scalar field and the electromagnetic kinetic term:\footnote{See Ref.~\cite{Olive:2001vz} for the most general action involving a scalar field, the Standard Model fields and the hypothetical DM particle.} 
\begin{eqnarray}
S  = \frac{1}{2} \int d^4x \sqrt{-g}R + 
\int d^4x \sqrt{-g} \left[   \frac{1}{2} \partial^{\mu}\varphi\partial_{\mu}\varphi - V(\varphi) \right] -  \frac{1}{4} \int d^4x \sqrt{-g}  B_F F_{\mu \nu}F^{\mu \nu}\;.
\end{eqnarray}
Varying with respect to $\varphi$ yields the following equation of motion:
\begin{equation}
	\square\varphi + V_\varphi + \frac{1}{4} B_{F\varphi} F_{\mu \nu}F^{\mu \nu} = 0\;,
\end{equation}
where $B_{F\varphi}$ is the $\varphi$ derivative of the coupling term. In principle, the electromagnetic field sources the scalar field, but it is important to note that the evolution of the scalar field does not depend on $B_F$ as the statistical average of $F^{\mu\nu}F_{\mu\nu}$ is zero.

Indeed the Lagrangian above allows us to define an ``effective'' fine-structure constant:
\begin{equation} \label{alpha}
\alpha(t)=\frac{\alpha_{0}}{B_F(\varphi(t))}
\end{equation}
where $\alpha_{0}$ is the value measured today. From \eqref{alpha}, we obtain the relative variation:
\begin{equation} \label{Dalpha}
\frac{\Delta \alpha}{\alpha}\equiv \frac{\alpha(t) - \alpha_{0}}{\alpha_{0}} = \frac{1}{B_F(\varphi(t))} - 1\,.
\end{equation}
It is evident that the possible evolution of the fine-structure constant depends on both the evolution of the scalar field and the coupling $B_F$.

Still following Ref.~\cite{Bekenstein:1982eu}, we  consider a simple exponential coupling:
\begin{equation} \label{beke}
B_F =  e^{- \zeta ( \varphi - \varphi|_{t_0} )} \simeq 1 -  \zeta \, \varphi - \varphi|_{t_0}\;,
\end{equation}
which satisfies the condition $B_F(\varphi(t_0))=1$ so that today we have $\alpha(t_0) = \alpha_0$.
The approximation comes from the fact that, see Ref.~\cite{Marra:2005yt},  the coupling constant $\zeta$ is constrained from observations to be small ($\zeta \sim 10^{-6}$). \

It is straightforward to generalize Eqs.~\eqref{alpha} and \eqref{Dalpha} if the scalar field has also a spatial dependence:
\begin{equation} \label{alphaMcV}
\alpha(t,r)=\frac{\alpha_{0}}{B_F(\varphi(t,r))}\; ,
\end{equation}
\begin{equation} \label{DalphaMcV}
\frac{\Delta \alpha}{\alpha}\equiv \frac{\alpha(t,r) - \alpha_{0}}{\alpha_{0}} = \frac{1}{B_F(\varphi(t,r))} - 1\; ,
\end{equation}
where $\alpha_0$ is a reference value. As long as the total variation of $\varphi$ is small we can also generalize Eq.~\eqref{beke}:
\begin{equation}\label{bekeMcV}
    B_F =  e^{- \zeta ( \varphi - \varphi|_{t_0,r_0} )} \simeq 1 -  \zeta \, \varphi\left(t,r\right) - \varphi|_{t_0,r_0}\;,
\end{equation}
where $t_0$ and $r_0$ stands for today and here.  Eqs.~\eqref{DalphaMcV} and \eqref{bekeMcV} can then be combined to give:
\begin{equation}\label{DalphaDE}
    \frac{\Delta \alpha}{\alpha} \simeq \zeta\delta\varphi(t,r) \; .
\end{equation}


\section{First-order perturbations in the Newtonian Gauge}\label{sec1}
We are interested in scalar perturbations around a FLRW background. Our perturbed metric, in the Newtonian gauge, is then:
\begin{equation}
    ds^2 = -(1+2\Psi)dt^2 + a^2(t)\delta_{ij}(1 +2\Phi)dx^idx^j \; .
\end{equation}
We consider for simplicity a matter sector being composed of a barotropic perfect fluid with energy momentum tensor given by:
\begin{equation}
    T_{m \nu}^{\mu} = \left(\bar{\rho} + P\right)u^{\mu}u_{\nu} + P\delta^{\mu}_{\nu} \; ,
\end{equation}
with:
\begin{eqnarray}
\bar{\rho} &=& \rho\left(1+\delta\right)\; ,\\
P &=& w\rho + p\; , \\ 
u^{\mu}&=& \left[1-\Psi, \frac{v_i}{a}\right]\; ,
\end{eqnarray}
where $\delta, p$ and $v_i$ are first-order quantities.
Finally, we are considering a Quintessence dark energy fluid with energy-momentum tensor: \begin{equation}
    T_{\bar{\varphi} \nu}^{\mu} = g^{\mu\beta}\partial_{\beta}\bar{\varphi}\partial_\nu\bar{\varphi} - \delta^{\mu}_{\nu}\left(\frac{1}{2}g^{\alpha\beta}\partial_\alpha \bar{\varphi}\partial_\beta \bar{\varphi} + V(\bar{\varphi})\right)\; ,
\end{equation}
where we decompose the scalar field as $\bar{\varphi} = \varphi + \delta\varphi$, with $\delta\varphi$ being a perturbation.
The resulting first-order Einstein field equations are, see for example Ref.~\cite{Piattella:2018hvi}:
\begin{equation}\label{frieq}
     -\frac{2}{a^2}\nabla^2 \Phi +6H\dot{\Phi} =\rho \delta + V_{,\varphi}\delta\varphi + \Psi\left(2V + 2\rho\right) + \dot{\varphi} \dot{\delta\varphi}\; ,
\end{equation}
\begin{equation}\label{acceq}
    2\left[\frac{1}{3a^2}\nabla^2\left(\Psi + \Phi\right) + H\left(\dot{\Psi} - 3\dot{\Phi}\right) - \ddot{\Phi}\right] = -2\Psi\left(V - w\rho\right)+ p - V_{,\varphi}\delta\varphi  + \dot{\varphi}\dot{\delta\varphi} \; ,
\end{equation}
\begin{equation}\label{0ieq}
    -av^i_{,i}\rho\left(1+w\right) + \dot{\varphi}\nabla^2\delta\varphi - 2\nabla^2\left(H\Psi - \dot\Phi\right)   = 0\;,
\end{equation}
while the Klein-Gordon and continuity equations, where we are considering negligible the contribution of the electromagnetic field,  become:
\begin{equation}\label{kggeneq}
    \ddot{\delta\varphi} + 3H\dot{\delta\varphi} -\frac{\nabla^2}{a^2}\delta\varphi + V_{,\varphi\varphi}\delta\varphi= -2\Psi V_{,\varphi} + \dot{\varphi}\left(\dot\Psi - 3\dot{\Phi}\right) \; , 
\end{equation}
\begin{equation}\label{densitycontreq}
  \dot{\delta} + 3H\delta\left(-w+ \frac{p}{\delta\rho}\right) + \left(1+w\right)\left(3\dot{\Phi} + \frac{v^{i}_{,i}}{a}\right)= 0 \;. 
\end{equation}
Finally, we have the Euler equation for the velocity field $\theta = \frac{v^i_{,i}}{a}$:
\begin{equation}\label{thetaeq}
    \dot{\theta} + 2H\theta = \frac{\nabla^2}{a^2}\Phi - \frac{\nabla^2}{a^2}\frac{p}{\rho\left(1+w\right)} \;. 
\end{equation}

\section{Spherically symmetric perturbations during slow roll}

Since we are interested in the effect of structures on a dynamical DE Quintessence field, we make the following assumptions:

\begin{itemize}
    \item We assume spherical symmetry, \textit{i.e.}, all perturbations will be function of time and radius only (while background quantities are of course function of time only).This simplifying assumption, at the cost of some generality, allows us to write down explicitly and in a simple way the Laplacian operator, allowing us for an analytical representation of the solutions.

    \item We assume that the scalar field $\varphi$ is in slow-roll and compatible with a DE behavior capable of mimicking a cosmological constant, with an equation of state parameter $w_{\varphi} \approx -1$. We can define the following slow roll parameters in terms of the potential of the scalar field:
    \begin{equation}\label{slowrollconditions}
        \epsilon_V \equiv \frac{1}{2}\left(\frac{V_{,\varphi}}{V}\right)^2 \; \qquad \eta_{V}\equiv \frac{1}{2}\frac{V_{,\varphi\varphi}}{V} \; ,
    \end{equation}
    and the hypothesis of slow roll implies $\epsilon_V, \eta_V \ll 1$, which we will consider as perturbative quantities. Slowly rolling DE is motivated by its similarity with the inflaton field for producing an accelerated expansion, and can be obtained in such a way that it does not alter the cosmological evolution at early times. Known potentials which allow such a behaviour for the scalar field can be generally classified into \textit{freezing} and \textit{thawing}  models, see for example chapter 7 of  Ref.~\cite{2010deto.book.....A}.

    \item We assume that $\rho$ is composed only of dust, whose perturbations are pressureless, \textit{i.e.}, $\rho \propto a^{-3}$ and $w,p = 0$.  Therefore, we are not including in our description of the inhomogeneities of the matter power spectrum the Maxwell field. This is because our main focus is to understand the impact of these inhomogeneities on the quintessence field during the late stages of the evolution of the Universe, when radiation has already diluted enough due to the Hubble flow. Therefore, the radiation energy density itself could be considered as a perturbation. As we mention in Section \ref{sec2}, the Quintessence field is in principle sourced by the Maxwell field, but at background level this can be neglected as in FLRW the average of $F_{\mu\nu}F^{\mu\nu}$ vanishes. On the other hand, one can argue that the very same coupling should in principle change the continuity equation of the radiation field, beyond the appearance of an effective non-constant $\alpha$ (where the latter can be defined  in terms of the DE field only, see for example Refs.~\cite{Martins:2015ama,PhysRevD.84.023518}).  However, this coupling is explicitly given in terms of the product of the radiation energy density and the time derivative of the scalar field, see for example Eq.~(7) of Ref.~\cite{Sandvik:2001rv} or Eq.~(19) of Ref.~\cite{Barrow:2013uza} . Since we are assuming that the scalar field is in slow roll, the coupling becomes then a purely second-order quantity, and can be safely neglected in our linear order analysis focused on the changes of the DE field.
\end{itemize}

\subsection{Static configurations}
The first case we are going to consider is a toy model in which all perturbations are assumed to be stationary or slowly varying, therefore we can neglect time derivatives of the scalar field perturbations with respect to spatial ones, and consider a static gravitational potential $\Phi = \Phi_S$.
Since the system is stationary, we also assume that the particles of dust are in orbital motion around the center of symmetry, \textit{i.e.} $\theta = 0$, and therefore there is no infalling or escaping matter. In this case, Eq.~\eqref{thetaeq} implies that $\Phi_S = \Phi_0/r$ and the remaining Einstein field equations become:
\begin{align}
    \rho\left(\delta - 2\Phi_S\right) &= 0\;, \\
    \label{deltaphistatic}
     2\Phi_S V &= \delta\varphi V_{,\varphi} \;,\\
     \label{laplvarphistatic}
     \nabla^2\delta\varphi &= 0 \;,
     \end{align}
     while the Klein-Gordon equation for the scalar field perturbation become:
     \begin{equation}
     \label{kgeqstatic}
    \frac{\nabla^2 \delta\varphi}{\delta\varphi} = V a^2\left(\frac{V_{,\varphi \varphi}}{V} - \frac{V_{,\varphi}^2}{V^2}\right) \quad \implies \quad \nabla^2\delta\varphi  = 2 V a^2 \left(\eta_V-\epsilon_V\right)\delta\varphi\; .
     \end{equation}
Note that Eqs.~\eqref{laplvarphistatic} and \eqref{kgeqstatic} are not in disagreement since $\nabla^2\delta\varphi$ is manifestly a second-order quantity during slow-roll. Therefore, we are left with the simple solution for $\delta\varphi$:
\begin{equation}
    \delta\varphi= \frac{F_1}{r}\;.
\end{equation}
The constant $F_1$ is fixed by imposing that is compatible with the remaining field equations, so Eq.~\eqref{deltaphistatic} gives us:
\begin{equation}
F_1 = 2\frac{V}{V_{,\varphi}}\Phi_0\; .
\end{equation}
Therefore, we have obtained that the gravitational field and the scalar field perturbations in a system described by a static density contrast are proportional to $\delta = \delta_0/r=2\Phi_0/r$.
This is not surprising, since the solution we obtained is the same than for a massless scalar field in a Schwarzschild background. Note that the perturbations are ``massless" only because the background field is in slow-roll, otherwise the Klein-Gordon equation would be solvable only for an exponential potential of the form $V(\varphi) = e^{\kappa\varphi}$.

\subsection{Non-static configurations}

Now, we want to relax the assumption of a static gravitational potential and allow for a non-vanishing velocity field ($\theta \neq 0$). On the other hand, since our main interest is to study how the Quintessence field is modified at linear level by the presence of a stable structure, we shall assume that the density profile of the structure does not change in time with respect to its surroundings. In other words, we now assume $\dot{\delta}\sim 0$.
Under this assumption, Eqs.~\eqref{densitycontreq} and \eqref{thetaeq} can be combined to obtain a closed differential equation for $\Phi$, indeed the former is simply:
\begin{equation}
\dot{\Phi}= -\frac{\theta}{3}\;,    
\end{equation}
therefore Eq.~\eqref{thetaeq} becomes:
\begin{equation}
    \label{nablaphi}
    \left(\ddot{\Phi} + 2H\dot{\Phi}\right)3a^2 = -\nabla^2\Phi \; .
\end{equation}
Assuming variable separation $\Phi= T(t) R(r)$, we have:
\begin{equation}
    -3a^2\left(\frac{\ddot{T}}{T}+2H\frac{\dot{T}}{T}\right)=\frac{\nabla^2 R}{R}=\kappa \;,
\end{equation}
with $\kappa$ a constant. For the radial part we have:
\begin{equation}
R(r) = \frac{c_1e^{-\sqrt{\kappa}r}}{r}\;.
\end{equation}
The time part can be solved analytically at the various stages of cosmological evolution. In particular, during the Matter and Dark Energy-dominated epochs, $a_{MD}(t) = a_*t^{2/3}$ and $a_{DE}= a_*e^{Ht}$ respectively, with $a_{\star}$ a reference scale factor. Defining $\lambda = \sqrt{\kappa/3a_*^2}$, the general solutions are:
\begin{equation}
    T(t) = \begin{cases} \dfrac{F_1 \cos{\left(3\lambda t^{\frac{1}{3}}\right)} + F_2\sin{\left(3\lambda t^{\frac{1}{3}}\right)}}{\lambda t^{\frac{1}{3}}} \; &\text{Matter-dominated epoch} \\[0.6em]
    D_1\dfrac{e^{-Ht}}{H}\lambda J_{1}\left(\frac{e^{-Ht}}{H}\lambda\right) \; &\text{Dark Energy-dominated epoch}
    \end{cases}
\end{equation}
We now consider the particular case of $\kappa = 0$, which recovers the gravitational potential $ \propto 1/r$ arising in the static configuration. In this case, the temporal part of the gravitational potential goes as:
\begin{equation}
    T(t) \propto \begin{cases}  a^{-\frac{1}{2}} \; &\text{Matter-dominated epoch}\\
    \frac{1}{H}a^{-2} \;&\text{DE-dominated epoch
    }
    \end{cases}
\end{equation}

We can now subtract Eqs.~\eqref{frieq} and \eqref{acceq} to obtain:
\begin{equation}
    \ddot{\Phi} + 7H\dot\Phi +  \Phi\left(2V+\rho\right) = \frac{\rho\delta}{2} + V_{,\varphi}\delta\varphi \; ,
\end{equation}
where we can use our solution for $\Phi$ to obtain:
\begin{equation}
    \delta\varphi = \begin{cases}   
    \dfrac{1}{V_{,\varphi}}\left[\Phi\left(\dfrac{H^2}{2} + \dfrac{5V}{4}\right) - \dfrac{\rho\delta}{2}\right] \; &\text{Matter-dominated epoch}\\[0.5cm]
    \dfrac{1}{V_{,\varphi}}\left[-4\Phi H^2 - \rho\left(\dfrac{\delta}{2} + \Phi\right)\right]\; &\text{Dark Energy-dominated epoch}
    \end{cases}
\end{equation}
Note that if in the Matter epoch we can neglect the potential $V$ with respect to the matter density $\rho$, and vice versa during the DE-dominated epoch, the above expressions simplify further and we have:
\begin{equation}\label{deltavarphisol}
    \delta\varphi = \begin{cases}
    \dfrac{\rho}{V_{,\varphi}}\left(\dfrac{\Phi}{6} - \dfrac{\delta}{2}\right) \; &\text{Matter-dominated epoch}\\[0.5cm]
  -\dfrac{4V}{3V_{,\varphi}}\Phi -\dfrac{\rho\delta}{2V_{,\varphi}} \; &\text{DE-dominated epoch}
    \end{cases}
\end{equation}
Note that the solutions of \eqref{deltavarphisol} are well-defined as long as the derivatives of $V$ are finite. This is however expected as $V_{,\varphi}$ is proportional to the slow-roll parameter $\epsilon_V$ defined in \eqref{slowrollconditions}, which must be small but not vanishing in order for DE to be dynamical and different from a cosmological constant $\Lambda$.
We see from this set of solutions that the spatial dependence of $\delta\varphi$ is given by a particular combination of $\delta$ and $\Phi$. So far, we have no information on the spatial dependence of the density contrast $\delta$, but under our assumption we know that it is constant with time. On the other hand, we know that in the static configuration the radial dependence of $\delta$ is proportional to the one of $\Phi$; if this is the case even for the non-static case, then $\delta\varphi$, $\Phi$ and $\delta$ all decay as $\propto 1/r$. The time part is, as expected, always decaying as structures are disrupted by the Hubble flow. The specific rate of decay depends on the form of the potential $V$, but we can nevertheless draw some conclusions simply by looking at Eq.~\eqref{deltavarphisol}. Indeed, we can imagine that towards the end of matter domination, since $\Phi$ is decaying whilst $\delta$ is constant in time, the leading contribution to $\delta\varphi$ goes as $\rho/V_{,\varphi}$. During DE domination, if $V \propto a^{-1}$, both the terms proportional to $\delta$ and $\Phi$ will have the same time dependence. If $V$ decays faster, then the main contribution will be given by the term proportional to $\delta$. If $V$ decays slower, it will be given by the term proportional to $\Phi$.

\subsection{Discussion}
Our main goal was to study the effects induced by the presence of stable structures on a dynamical DE field represented by a Quintessence field in slow-roll. 
To model a stable structure, we have considered a static, spherically symmetric distribution for the density contrast $\delta$. 
Since $\delta = \delta\rho/\rho$, we are therefore assuming that the matter perturbations $\delta\rho$ have the same time dependence as the background matter density $\rho$.

Within the above assumptions, we were able to find analytical solutions for the full system of Einstein field equations assuming firstly a static, stationary configuration, and then a non-static one compatible with $\dot{\delta}=0$.
If $\delta$ and $\Phi$ have the same spatial dependence, the latter is inherited by the scalar field perturbations $\delta\varphi$. Since in the static case we have found $\delta, \Phi \propto 1/r$, we expect the same to be true in the non-static case and therefore $\delta\varphi \propto 1/r$.
The time evolution of $\delta\varphi$ depends instead on the specific stage of the evolution of the Universe in which these perturbations are generated. Once the form of the scalar field potential $V$ is specified, one can solve the background Klein-Gordon equation to obtain $\varphi$ (and then $V$) as a function of time, then substitute in Eq.~\eqref{deltavarphisol} for $\delta\varphi$.
Our main result, obtained by directly solving the perturbed EFE, is that due to the presence of a gravitational potential generated by a stable structure, $\delta\varphi$ depends on the radius as $\delta\varphi \propto 1/r$.\footnote{In the most general case, $\propto e^{-\kappa r}/r$, but as we saw we should consider $\kappa = 0$ if we want to recover the static configuration for $\theta = 0$.}

\begin{figure}
    \centering
    \includegraphics[scale=1.2]{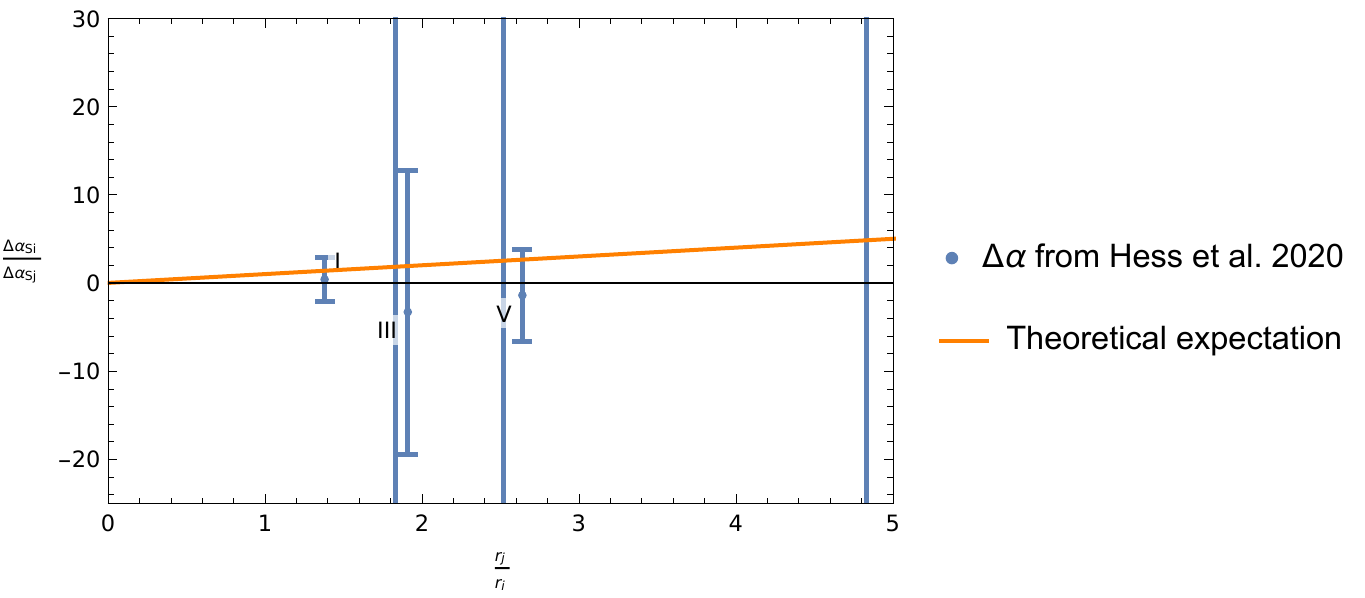}
    \caption{Ratios between the relative variation of $\alpha$ for different pairs of stars together with the theoretical expectation, \textit{i.e.} the bisector. We did not include measurements involving S0-13 as the resulting error bars are orders of magnitude bigger than they central value. The solid orange line represents the theoretical prediction $\Delta\alpha_i/\Delta\alpha_j = r_j/r_i$.}
    \label{Plot}
\end{figure}

Assuming a coupling \textit{à la} Bekenstein, the spatial dependence of the scalar field is  inherited by  $B_F(\varphi)$ and therefore by $\alpha$. In Ref.~\cite{Hees:2020gda}, constraints on the variation of $\alpha$ were obtained using spectroscopic measurements of 5 stars orbiting around the supermassive black hole in the center of the Milky Way. Our results predict that the ratio of the $ \Delta \alpha$ variations between pairs of stars are inversely proportional to their distance, \textit{i.e.}, for two stars $A$ and $B$, we have $\Delta\alpha_A/\Delta\alpha_B \simeq r_B/r_A$. In Table \ref{liststars}, we report the observed values of $\Delta \alpha$ for each star from Ref.~\cite{Hees:2020gda}, complemented with their distances in arcsec from the central black hole taken from Ref.~\cite{2013ApJ...764..154D}. The results of our prediction against the observations are reported in Table \ref{listresults} and Fig.\ref{Plot}.
\begin{table}[h!]
    \centering
    \begin{tabular}{c||c|c|}
          Star \; \;\; & $\;\;\Delta\alpha_{star}/\alpha_{ref}\;\;$ & $\;\;r_{star} \left[arcsec\right]\;\;$ \\
         \hline
          S0-6 & $\;1.0 \pm 1.2 \;\times 10^{-4}$ & $0.36 $\\
            S0-12 & $-0.3 \pm 1.4 \times 10^{-4}$ & $0.69 $\\
              S0-13 & $\;0.03 \pm 3.5\times 10^{-4}$ & $0.69$\\
                S1-5 & $-0.7 \pm 2.4 \times 10^{-4}$ & $0.95 $\\
                  S1-23 & $\;0.9 \pm 5.8\; \times 10^{-6}$ & $1.74$
    \end{tabular}  
    
    \caption{Measurements of $\Delta \alpha$ from Ref.~\cite{Hees:2020gda} through observations of 5 stars orbiting around Sagittarius A*. Their distances $r$ from the black hole, in arcseconds, are taken from Ref.~\cite{2013ApJ...764..154D}. }
    \label{liststars}
\end{table}

\begin{table}[h!]
    \centering
    \begin{tabular}{c||c|c|}
          Pair of Stars $S_{i},S_j$\; \;\; & $\;\;\Delta\alpha_{Si}/\Delta\alpha_{Sj}\;\;$ & $\;\;r_{j}/r_{i} \;\;$ \\
         \hline
          I (S0-12,S1-5) &$0.4 \pm 2.5$& $1.38$\\
          II (S1-5,S1-23)&$-77.8 \pm567.9$&1.83\\
          III (S0-6,S0-12) & $\;-3.3 \pm 16.0 \;$ & $1.91 $\\
         IV (S0-12,S1-23)&$-33.3 \pm 264.9$& $2.52$ \\
             V (S0-6,S1-5) & $\;-1.4 \pm 5.2$ & $2.64$\\
                VI (S0-6,S1-23) &  $\;111.1 \pm 728.4\;$ & $4.83$
    \end{tabular}  
    
    \caption{Ratios between pairs of $\Delta \alpha$ measurements from Ref.~\cite{Hees:2020gda} together with the inverse ratio of their distances from the Black Hole. We did not include pairs containing the star S0-13 as the resulting error bars are orders of magnitude bigger than their central values.  }
    \label{listresults}
\end{table}

If, in the future, the number and the precision of this kind of spectroscopic observations improve and variations of $\alpha$ are detected (or not) with enough statistical accuracy, then these electromagnetic signals will allow to constrain the nature of Dark Energy. The results obtained in this work suggest that spectroscopic measurements offer a new  way to search for Dark Energy observables through the spatial variations of the fine-structure constant. This framework is especially interesting because it predicts, even for choices of the  potential which make the Quintessence field indistinguishable from a cosmological constant at the background level, a measurable deviation at linear order in perturbation on astrophysical scales.
To explore further this possibility, we could relax the slow-roll hypothesis, or we might as well derive predictions in other Dark Energy scenarios. We leave these compelling directions for future works.

\begin{acknowledgments}
The authors acknowledge the \textit{Funda\c{c}\~ao de Amparo \`a Pesquisa e Inova\c{c}\~ao do Esp\'irito Santo} (FAPES), the \textit{Conselho Nacional de Desenvolvimento Cient\'ifico e Tecnol\'ogico} (CNPq) and the  \textit{Coordena\c{c}\~ao de Aperfei\c{c}oamento de Pessoal de N\'ivel Superior} (CAPES) for partial financial support. EF wishes to thank the Helsinki Institute of Physics for its kind hospitality. LG is grateful to the  ITP-Heidelberg, where part of this work was developed, for its kind hospitality
\end{acknowledgments}

\bibliography{references.bib}
\end{document}